\documentclass[%
 reprint,
 amsmath,amssymb,
 aps,
]{revtex4-2}

\usepackage{graphicx}
\usepackage{dcolumn}
\usepackage{bm}
\usepackage{xcolor}

\begin{document}

\preprint{APS/123-QED}

\title{Intrinsic Meron Spin Textures in Generic Focused Fields}

\author{Di Liu}
\author{Han Liu}
\author{Zheng Xi}%
 \email{zheng.xi@ustc.edu.cn}
\affiliation{%
 Department of Optics and Optical Engineering, University of Science and Technology of China}%




\date{\today}

\begin{abstract}
Optical spin textures with nontrivial topology hold promise for structured light and photonic information processing, yet their generation typically relies heavily on externally structured light with care. This raises questions about their universal existence and true robustness. Here, we uncover and experimentally verify a meron-like spin texture that emerges intrinsically in focused fields, without any wavefront engineering. This intrinsic meron spin texture, unlike their externally engineered counterparts, exhibits exceptional robustness against a wide range of inputs, including partially polarized and spatially disordered pupils corrupted by decoherence and depolarization. We attribute its resilience to topological protection from phase vortices in the focal field. Our findings reveal a naturally occurring spin structure that is intrinsic to the focused field with exceptional robustness against noise, which complements the existing externally engineered ones. It offers new ingredients into topological spin textures in optics and enriches their potentials for disorder-resilient photonic applications.
\end{abstract}

\maketitle

Topological spin textures have attracted considerable attention in many fields across physics\cite{al2001skyrmions,bogdanov2003skyrmions,sondhi1993skyrmions,mühlbauer2009skyrmion,dzyaloshinsky1958thermodynamic,wang2025topological,ge2021observation}. In particular, driven by recent advances in structured light and engineered nanostructures, skyrmion-like configurations such as skyrmions, merons, bimerons, and hopfions have gained significant interest in optics\cite{tsesses2018optical,davis2020ultrafast,wan2022scalar,dreher2024spatiotemporal,shen2024optical,ehrmanntraut2023optical,shen2021topological,wang2024topological1,shen2022generation,du2019deep,he2024optical}. Although these textures are known for their robustness against perturbations, their realization in optical systems typically requires intricate setups and fine-tuned conditions. This stands in contrast to magnetism\cite{mühlbauer2009skyrmion,dzyaloshinsky1958thermodynamic,moriya1960anisotropic,finocchio2016magnetic}, where such textures can arise spontaneously from intrinsic material interactions, and their observations in most materials require minimal external engineering. This apparent discrepancy challenges the very notion of topological robustness and its universal existence in optics: if a spin texture is truly robust, it should emerge naturally under generic conditions. To date, however, no such intrinsically emergent spin texture has been conclusively identified in optical systems.

In this Letter, we resolve this discrepancy and show that the most generic spin texture in focused optical fields is a meron-like spin texture that arises spontaneously without engineered input. Consequently, it exhibits exceptional robustness across a wide range of conditions. By varying the input polarization over the entire Poincaré sphere, we demonstrate that the meron pattern consistently forms whenever the input polarization lies within either hemisphere, including partially polarized states corrupted by depolarization. Remarkably, the meron texture remains stable even under strong spatial randomness in the input polarization, amplitude, and phase. This level of resilience far exceeds that of previously reported externally engineered topological spin textures in optics. We attribute this to the robustness of phase vortices naturally occurring in the electromagnetic field components. Our work uncovers a
naturally occurring, topologically protected spin texture in optics with exceptional robustness that is of fundamental interest to a deeper understanding of topological robustness in optics and its applications. 

\begin{figure}[h]
    \includegraphics[width=\linewidth]{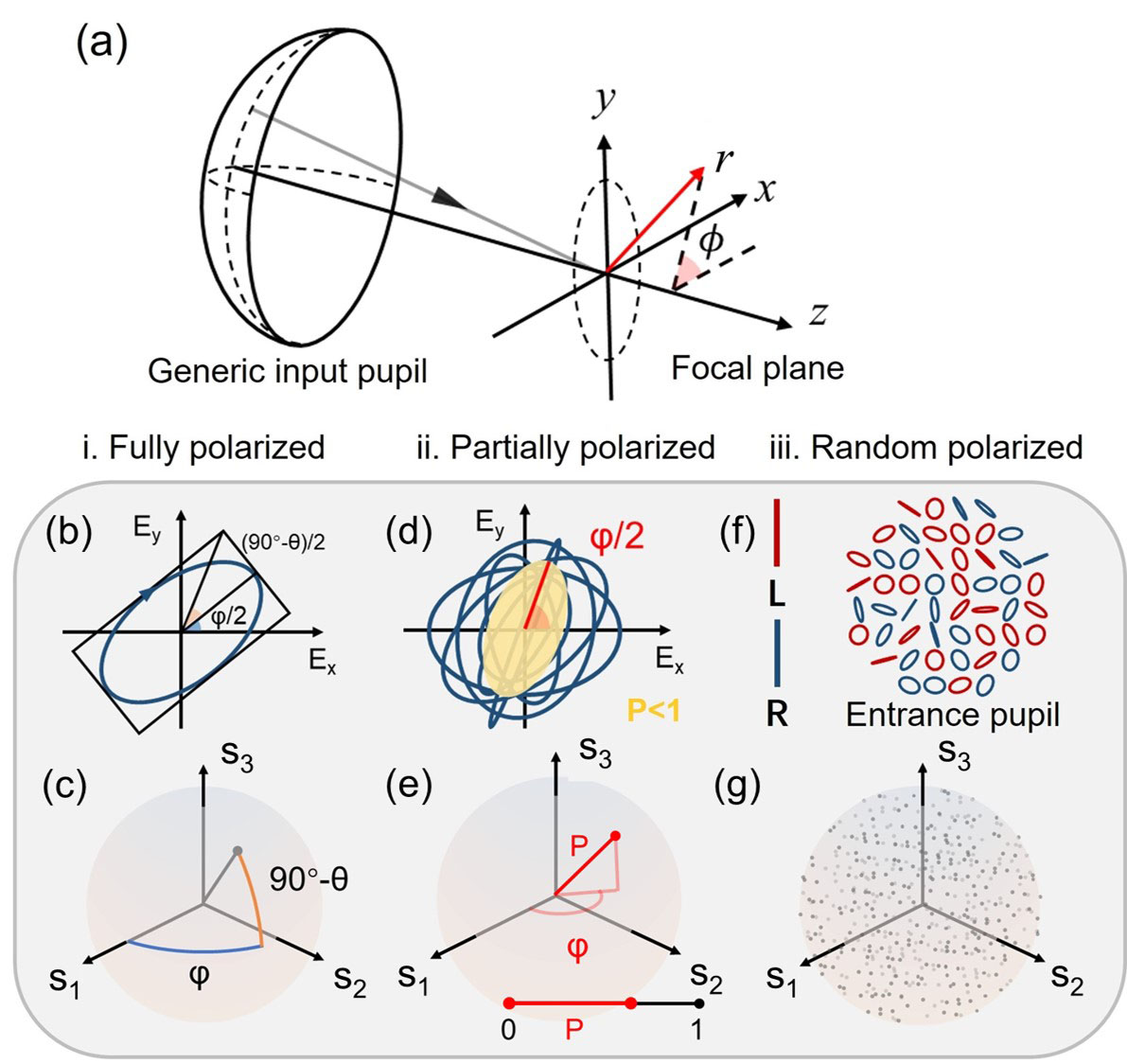}
    \caption{
 (a) The system under study, where the spin texture at the focal plane is examined with a generic input pupil. (b-g) are schematic diagrams of three input pupils under consideration: (b,c) Fully polarized, (d,e) Partially polarized, (f,g) Spatially-randomly polarized. }
    \label{fig:my_label}
\end{figure}

The system under study is shown in Fig. 1(a), where a pupil field is focused and the spin texture is examined at the focal plane. To capture the most generic spin textures, unlike previous works with engineered input, we apply no restrictions to the input polarization, allowing it to vary freely. 

Specifically, three different cases are considered as shown in Fig. 1(b-g): (i) a spatially uniform but arbitrarily polarized input pupil that is stationary in time (Fig. 1(b)). It corresponds to an arbitrarily chosen point on the surface of the Poincaré sphere in Fig. 1(c) with varying state of polarization. (ii) a spatially uniform but partially polarized state with polarization fluctuations in time (Fig. 1(d) where the blue ellipses are instant polarization states and the yellow one represents their ensemble averaging). This corresponds to an arbitrarily chosen point within the Poincaré sphere with degree of polarization $P$ less than one (Fig. 1(e)). (iii) a pupil with spatially varying polarizations(Fig. 1(f)), with 
spatial polarization distributions covering part of the Poincaré sphere (Fig. 1(g)). These three cases cover a broad range of input field and allow us to observe the most generic spin textures in a focused field.

We begin with a spatially uniform pupil, including cases (i) and (ii). These inputs span the full solid Poincaré sphere and can be described conveniently using the input polarization matrix $\Phi_{\text{in}}$ at the input pupil \cite{mandel1995optical}:
\begin{equation}
\Phi_{\text{in}} = P\Phi_p + (1 - P) \Phi_u,
\end{equation}
where $\Phi_p = |\psi\rangle_p \, {}_p\langle \psi|$ describes the polarized input pupil with $|\psi\rangle_p = \sin \frac{\theta}{2} \mathbf{e}_- + \cos \frac{\theta}{2} \mathbf{e}_+ e^{i\varphi}$ representing a point ($\theta$, $\varphi$) on the surface of Poincaré sphere under circularly polarized (CP) basis $\mathbf{e}_{+/-}=\frac{1}{\sqrt{2}}[1;\pm i]$. $\Phi_u$ is the $2\times2$ identity matrix representing  unpolarized pupil. By varying $\theta$, $\varphi$, and $P$ (degree of polarization), Eq. (1) covers all possible polarization states in the full solid 
Poincaré sphere.

The local spin density $\mathbf{s}$ at the focal plane $z=0$ from this generic pupil can be obtained analytically using the Richards–Wolf diffraction theory\cite{wolf1959electromagnetic,supp}:
\begin{equation}
\mathbf{s} = \begin{bmatrix}
s_r \\
s_\phi \\
s_z
\end{bmatrix} = \frac{k^2 f^2}{4 I(r)} \begin{bmatrix}
2 \mathcal{I}_1 (\mathcal{I}_0 - \mathcal{I}_2) P \sin \theta \sin (2\phi-\varphi) \\
2 \mathcal{I}_1 (\mathcal{I}_0 + \mathcal{I}_2) [1 + P \sin \theta \cos (2\phi-\varphi)] \\
(\mathcal{I}_0 + \mathcal{I}_2) (\mathcal{I}_0 - \mathcal{I}_2) P \cos \theta
\end{bmatrix},
\end{equation}
where $k$ is the wavenumber, $f$ is the focal length and $I(r)$ is the focal intensity at position $r$ at the focal plane with $z=0$. $\mathcal{I}_0$, $\mathcal{I}_1$, $\mathcal{I}_2$ are the three  diffraction integrals.

Eq. (2) reveals a remarkably robust spin texture: whenever the input pupil carries nonzero net spin with $P\neq0$ and $\theta\neq\pi/2$, the resulting spin texture at the focal plane always exhibits a purely longitudinal spin $s_z$ at the center (where $\mathcal{I}_1=\mathcal{I}_2=0$ and $\mathcal{I}_0\neq0$ ), surrounded by a ring of purely azimuthal transverse spin $s_\phi$ located at the boundary defined by $\mathcal{I}_0=\mathcal{I}_2$. This spin texture arises naturally from the focusing process and is largely independent of the specific input polarization state or degree, as long as the input pupil has some net spin. It thus represents a generic spin texture of focused fields.

One can compute the skyrmion number of this generic texture using
\begin{equation}
N_{\text{sk}} = \frac{1}{4\pi} \iint_\sigma \hat{\mathbf{n}} \cdot \left( \frac{\partial \hat{\mathbf{n}}}{\partial r} \times \frac{1}{r}\frac{\partial \hat{\mathbf{n}}}{\partial \phi} \right) r \text{d}r \text{d}\phi,
\end{equation}
where $\hat{\mathbf{n}}=\frac{\mathbf{s}}{|\mathbf{s}|}$ is the normalized spin vector, and the integral range $\sigma$ is the area within the boundary defined by $\mathcal{I}_0=\mathcal{I}_2$. As long as $P\cos\theta\neq0$, it takes half-integer value $N_{sk}=\pm0.5$ depending on the handedness of the input spin, which means that it is a meron texture. This skyrmion number can be intuitively understood as the product $N_{sk}=p\cdot m$, where $p=\pm0.5$ denotes the spin polarity (set by the sign of $s_z$ at the center)
, and $m=1$ is the vorticity of the in-plane transverse spin. The latter reflects the winding of the transverse spin around the azimuthal direction. The azimuthal spin boundary defined by $\mathcal{I}_0=\mathcal{I}_2$ acts as a natural disc boundary enclosing the topological structure. According to the Poincaré–Hopf theorem, $m$ equals the Euler characteristic of the disc, which is +1\cite{hopf2013differential,gbur2016singular}. Therefore $N_{sk}=p \cdot m = \pm 0.5$. 

\begin{figure}[h]
    \includegraphics[width=\linewidth]{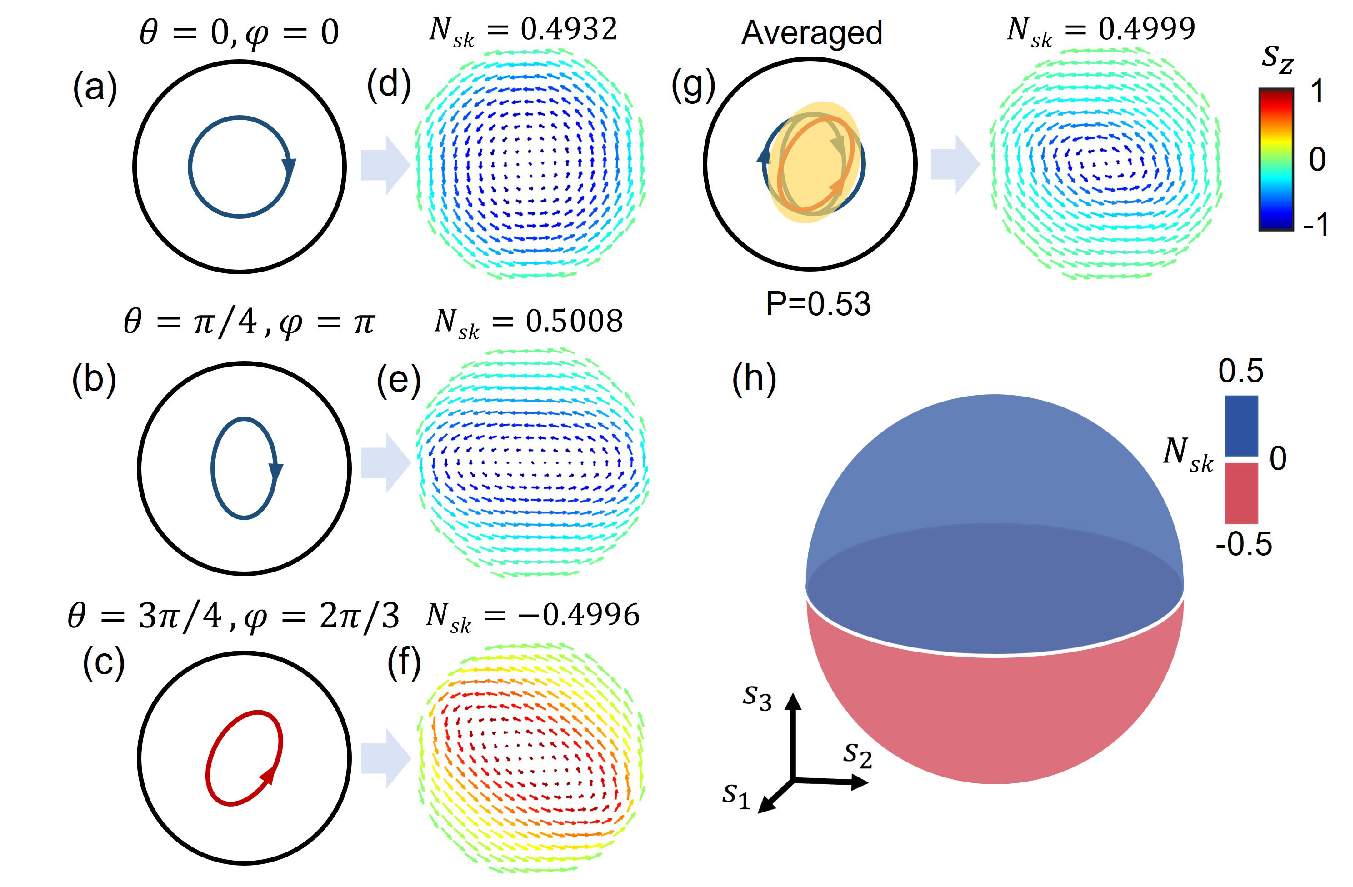}
    \caption{
Spin textures and skyrmion numbers under different incident conditions. (a-c) The incident polarization ellipse and (d-f) the normalized spin textures on the focal plane are shown. (g) The partially polarized state (yellow ellipse) formed by time averaging of the fully polarized state in (a-c) and the corresponding spin texture at the focal plane. (h) The solid Poincaré sphere is naturally divided into two halves with $N_{sk}=\pm0.5$.}
    \label{fig:my_label}
\end{figure}

To better visualize this spin texture, we present some examples under various input polarization states in Fig. 2(a-c). While the detailed spin density profiles at the focal plane may differ (Fig. 2(d-f)), the underlying skymion number remains unchanged with $N_{sk}=\pm0.5$, even for partially polarized inputs realized by time-averaging over randomly varying pure polarization states (Fig. 2(g))\cite{supp}.  

More generally, for a spatially uniform input pupil described by points on the solid Poincaré sphere, including both the surface and its interior using Eq. (1). The solid Poincaré sphere is naturally divided into two hemispheres by the condition $P\cos\theta=0$ as shown in Fig. 2(h). Whenever the input polarization lies within either hemisphere, corresponding to a field with net longitudinal spin, the resulting focal spin texture consistently exhibits the meron configuration with $N_{sk}=\pm0.5$. This reinforces its status as a fundamental and universal structure in focused optical fields.

One may ask what underlies the remarkable robustness of the meron spin texture. The central longitudinal spin component $s_z$ can be readily understood as arising directly from the net spin of the incident field. Specifically, when the input polarization lies on the equatorial plane of the Poincaré sphere when $P\cos\theta=0$, the net longitudinal spin vanishes, and correspondingly, the $s_z$ component at the focal center disappears.

More intriguing is the persistence of the robust transverse spin boundary characterized by a purely azimuthal component $s_\phi$. As shown in Eq. (2), this boundary exists even when the input pupil has no spin ($P=0$), which is completely different from  the longitudinal spin.

\begin{figure}[h]
    \includegraphics[width=\linewidth]{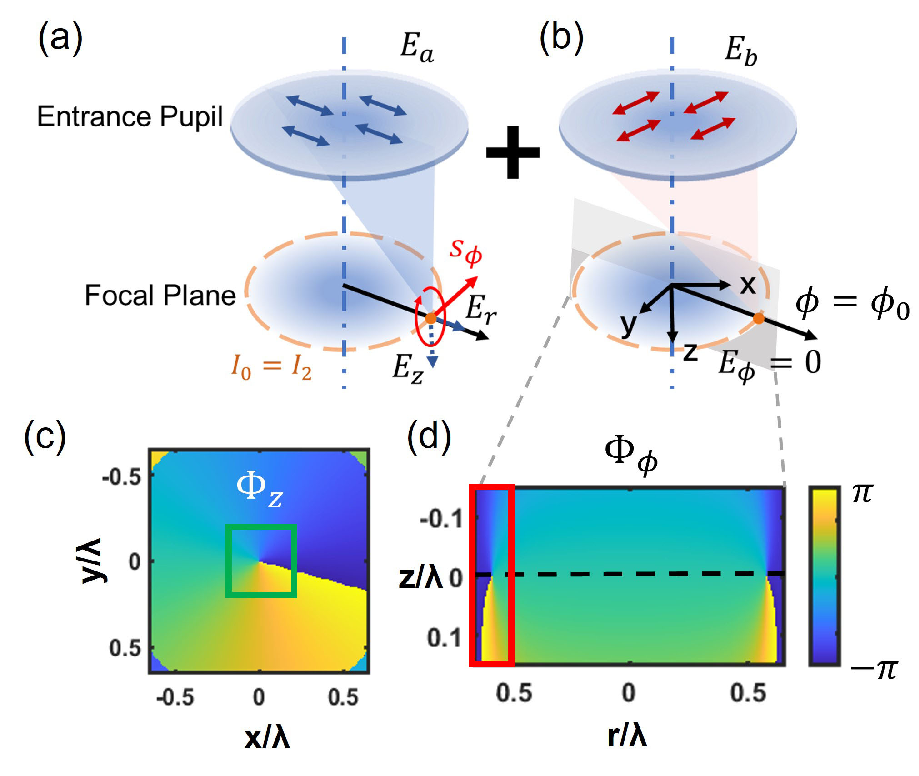}
    \caption{(a-b)  Decomposition of incident light into two orthogonal linearly polarized components. The orange dashed circle and orange dot represent the positions on the focal plane satisfy $\mathcal{I}_0=\mathcal{I}_2$. (c) The phase of $E_z$ in the $xy$ plane. (d) The phase of $E_\phi$ in the $rz$ plane. 
 }
    \label{fig:my_label}
\end{figure}

To gain more insight, we analyze the origin of the local transverse spin $s_\phi$ along an arbitrary azimuthal direction ($\phi=\phi_0$) in the focal plane. According to Eq. (2),$s_\phi=\frac{k^2f^2}{4I(r)}2 \mathcal{I}_1 (\mathcal{I}_0 + \mathcal{I}_2) [1 + P \sin \theta \cos (2\phi_0-\varphi)]$.
We decompose the input polarization into two orthogonal linear polarization components $E_a$ and $E_b$ as shown in Fig. 3(a) and 3(b) with $E_a=\cos\phi_0E_x+\sin\phi_0E_y$ aligned tangentially along $\phi_0$, and $E_b=-\sin\phi_0E_x+\cos\phi_0E_y$ perpendicular to it. $E_x$ and $E_y$ are the Cartesian pupil components as can be obtained from Eq. (1). This local basis allows us to capture how different polarization components contribute to the spin structure upon focusing.

Upon focusing, the two linear polarized pupil components $E_a$ and $E_b$ project differently into the focal field. The tangential $E_a$ component yields both the radial component $E_r=-\frac{ikf}{2}(\mathcal{I}_0+\mathcal{I}_2)E_a$ and the longitudinal $E_z=-\frac{kf}{2}2\mathcal{I}_1E_a$ at the focal plane as shown in Fig. 3(a), while $E_b$ gives rise to only $E_\phi=-\frac{ikf}{2}(\mathcal{I}_0-\mathcal{I}_2)E_b$. At the meron boundary defined by $\mathcal{I}_0=\mathcal{I}_2$ (orange dashed circle), the transverse field $E_\phi$ vanishes (Fig. 3(b)), and the local spin comes solely from $E_a$ with $\mathbf{s}=-\frac{1}{I(r)}Im\{E_r^*E_z\} \hat{\phi}=\frac{k^2f^2}{4I(r)}(\mathcal{I}_0+\mathcal{I}_2)2\mathcal{I}_1(1+P\sin\theta\cos(2\phi_0-\varphi))\hat{\phi}$, which is in agreement with Eq. (2). 

The above pupil decomposition reveals that the transverse spin $s_{\phi}$ along an arbitrary azimuthal direction at the meron boundary does not originate from circularly polarized input, but rather emerges geometrically from the projection of the linearly polarized component $E_a$ along this direction in the input pupil. Remarkably, at this meron boundary, the focal field is fully polarized with pure transverse spin $s_{\phi}$ even when the input pupil is partially or unpolarized with $P<1$ in Eq. (1)\cite{lindfors2007local,lindfors2005degree}. In fact, most input pupils can have nonzero decomposition of $E_a$ along all azimuthal directions $\phi_0$, even under unpolarized illumination where no phase correlation is present at the pupil. This explains the robustness of the meron boundary against depolarization, and the boundary even exists under unpolarized illumination with $P=0$\cite{supp}.

From the above analysis, it is clear that the presence of longitudinal and transverse spin components is guaranteed by the net longitudinal spin and the distribution of linearly polarized components across all azimuthal directions in the input pupil. However, their mere existence does not ensure a pure meron spin texture characterized by longitudinal spin $s_z$ at the center and purely azimuthal spin $s_\phi$ at the boundary. Two additional conditions must be satisfied: the transverse spin must vanish at the center ($s_\perp=0$), and the longitudinal spin must vanish at the boundary ($s_z=0$). These conditions are closely linked to the presence of phase vortices in the focal field components.

\begin{figure}[h]
    \includegraphics[width=\linewidth]{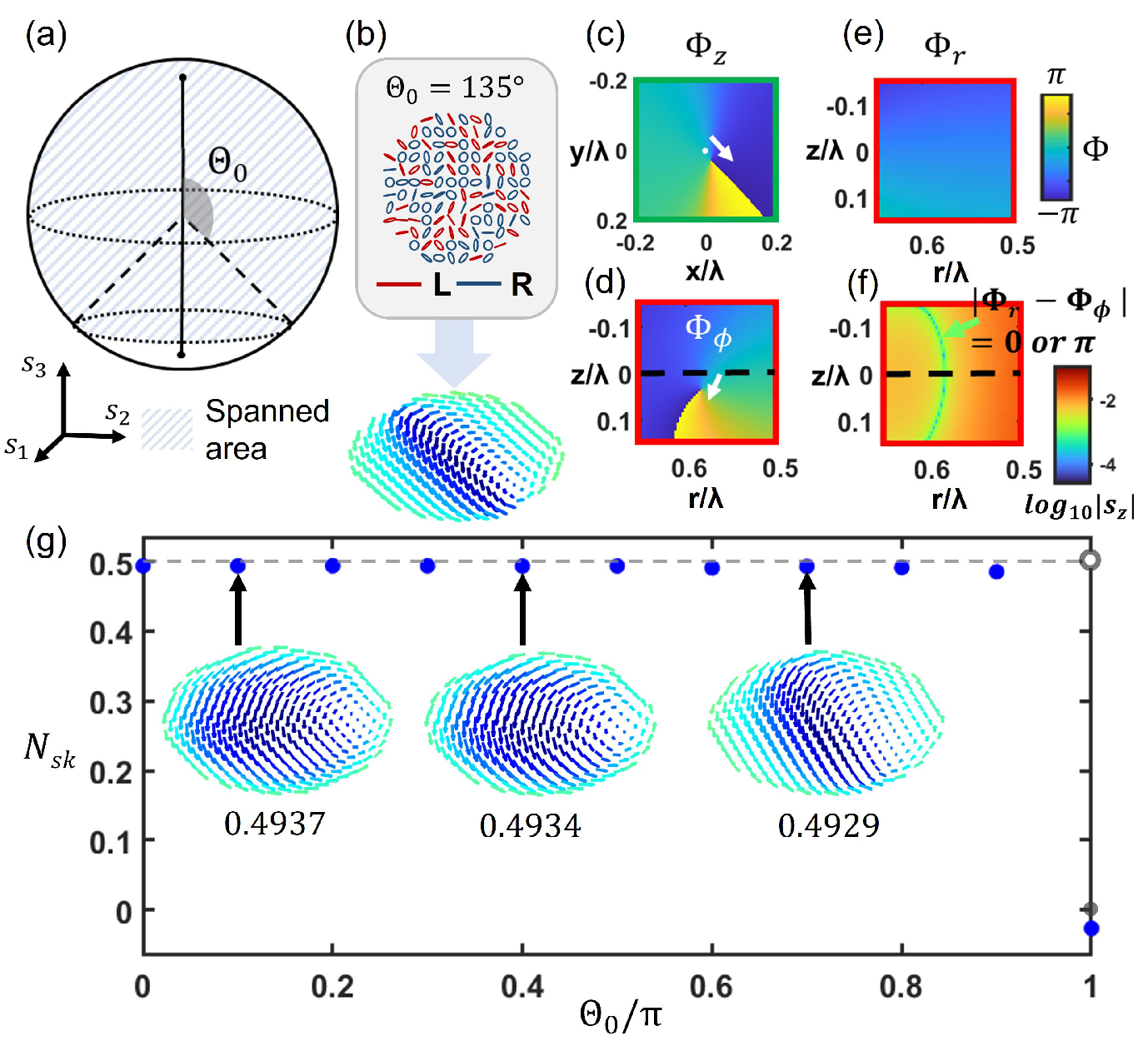}
    \caption{(a) Degree of spatial polarization randomness spanned by $\Theta_0$. (b) The polarization distribution at the input pupil when $\Theta_0=135^\circ$ and the corresponding spin texture at the focal plane. (c-e) Phase of $E_z$, $E_{\phi}$, and $E_{r}$ under the input pupil in (b). (f) $log_{10}|s_z|$ and the contour line in green highlights the positions with $|\Phi_r-\Phi_\phi|=0$ or $\pi$ (zero of $s_z$ with pure $s_\perp$). (g) Stability of $N_{sk}$ with increasing spatial randomness. The dashed gray line displays the theoretical values. The insets show the spin textures at different $\Theta_0$ with $N_{sk}$.
}
    \label{fig:my_label}
\end{figure}

At the focal point, the longitudinal field $E_z$ exhibits a phase vortex, as shown in Fig. 3(c), which originates from spin-orbit interaction inherent to focused systems\cite{zhao2007spin}. This vortex structure enforces $E_z=0$ at the center, leading to a vanishing $s_\perp\propto -Im({E_{r}}^*{E_z})\hat{\phi} + Im({E_{\phi}}^*{E_z})\hat{r}$, and thus a purely longitudinal spin $s_z$. Similarly, at the boundary defined by $\mathcal{I}_0=\mathcal{I}_2$, the azimuthal component $E_\phi=0$, and it possesses a phase vortex in the $rz$-plane (Fig.3(d)), causing $s_z\propto Im({E_r}^*{E_\phi})=0$ with a purely transverse spin $s_\phi$. These vortex structures in different field components ensure the existence of the pure longitudinal spin and transverse spin, and are essential to the robustness of the meron spin texture.

It is well known that phase vortex structures are topologically protected: under perturbations, their positions may shift continuously, but the vortices themselves cannot vanish. This intrinsic stability ensures that the underlying spin vortices or the meron spin texture remain robust even when the input field exhibits strong spatial randomness in the input pupil. 

To demonstrate this, we extend our analysis from cases (i) and (ii) to case (iii), where the input pupil exhibits a spatially varying polarization distribution.
By assigning  random values to $\theta$ and $\varphi$ within the ranges $\theta\in [0, \Theta_0]$ and $\phi\in[0, 2\pi]$ spatially across the pupil in Eq. (1) with P=1, we introduce spatial variations in amplitude and phase of different CP components, therefore generating an input pupil with random spatial polarizations\cite{supp}. By increasing $\Theta_0$, the area spanned by the spatially random polarization states on the Poincaré sphere is increased as shown in Fig. 4(a). This allows us to probe the stability of the meron texture under increasing degrees of spatial randomness in input polarization distributions.

As an example, we take $\Theta_0=135^\circ$ with the input pupil polarization distributions shown in Fig. 4(b) and the meron spin texture at the focal plane persists. As expected, the phase vortex in $E_z$ and $E_\phi$ do not vanish but only shift in position as can be seen in Fig. 4(c-d). The displacement of the central phase vortex in Fig. 4(c) results in a shift of the meron core with respect to the original focal point indicated by the white dot. The behavior at the boundary, however, requires more careful analysis. In the unperturbed case, the meron boundary is defined by the zero of the $E_\phi$ component, leading to $s_z=0$ and pure $s_{\phi}$. When spatial randomness is introduced, the phase vortex in $E_\phi$ moves within the $rz$-plane (Fig. 4(d)), and $E_\phi$ is not zero at the focal plane as for the unperturbed case. Although the vortex core in $E_\phi$ is no longer precisely located at the focal plane, as shown in Fig. 4(d) by the dashed line, it continues to induce rapid phase variations in its vicinity. The rapid phase variation in $E_\phi$ guarantees the in or out of phase relations with the other in-plane $E_r$ component, whose phase is nearly constant in the $rz$-plane(Fig. 4(e)). Together, they generate a curve of in-plane linear polarization in the focal region (highlighted curve in Fig. 4(f)) with $s_z=0$ and nonzero $s_{\phi}$ and $s_{r}$. This curve intersects with the focal plane (dashed line), and thus preserves the meron boundary.

\begin{figure}[h]    \includegraphics[width=\linewidth]{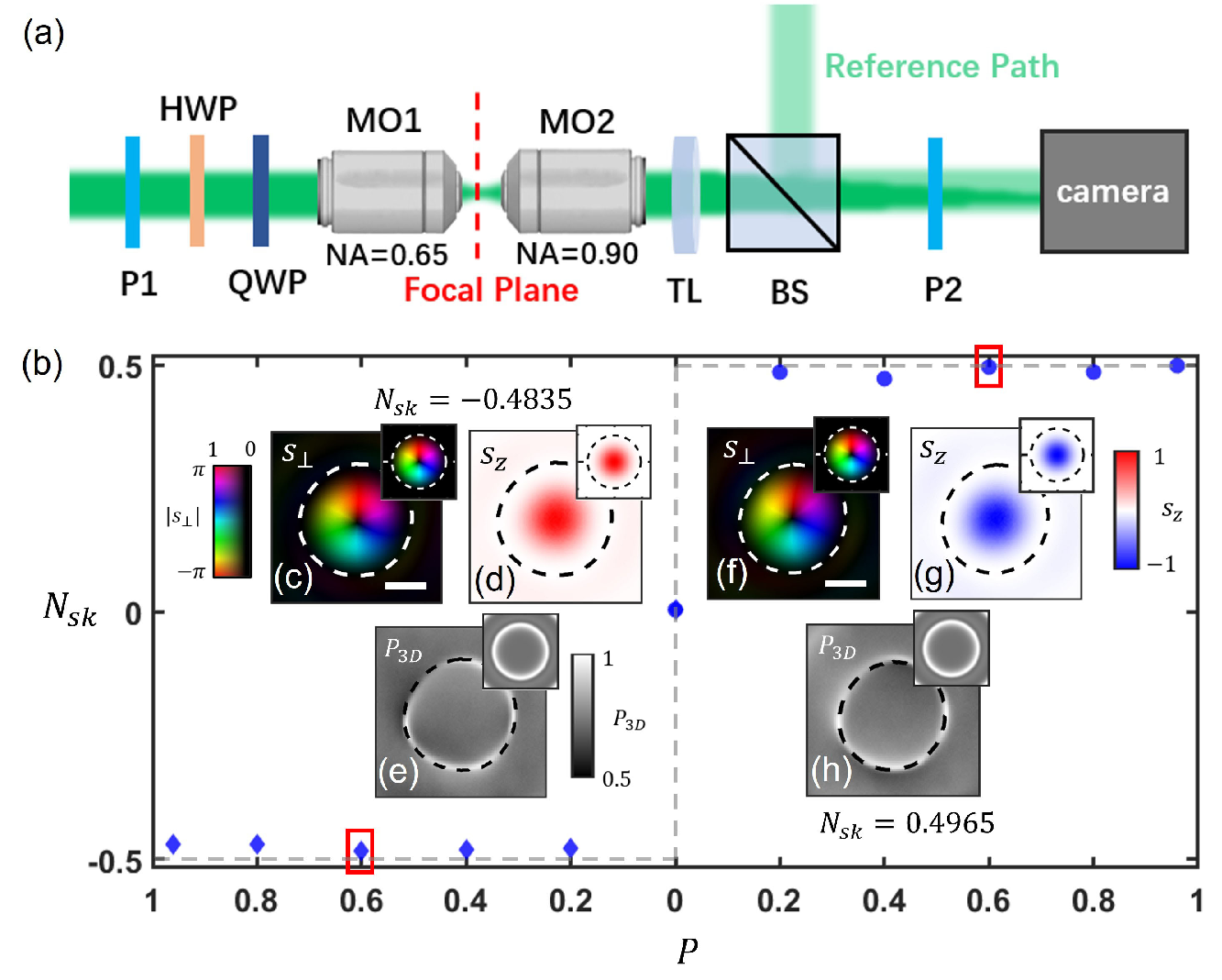}
    \caption{(a) Experimental setup based on off-axis holography for measuring spin meron structure in the focal region. P, polarizer; HWP, half-wave plate; QWP, quarter-wave plate; MO, microscope objective; TL, tube lens; BS, beam splitter. (b) Stability of $N_{sk}$ as a function of the input degree of polarization $P$. Insets: Detailed distributions at the states marked by the blue dots in the red boxes, showing (c)(f)the transverse and (d)(g)longitudinal spin components and (e)(h)the $P_{3D}$ distribution at the focal plane. In the transverse spin ($s_\perp$) map, color and brightness represent its direction ($\tan^{-1}(s_y/s_x)$) and magnitude, respectively. Theoretical simulations are shown for comparison. Scale bars, 500 nm.
}
    \label{fig:my_label}
\end{figure}

In Fig. 4(g), we summarize the outcomes for spatially perturbed input pupils with increasing spatial polarization randomness from 
$\Theta_0=0$, a homogeneous CP pupil to $\Theta_0=\pi$, a pupil with polarization state distributions covering the 
Poincaré sphere spatially. As long as the input field carries a net spin across the pupil with $\Theta_0<\pi$, corresponding to a spatial polarization distribution that does not fully span the Poincaré sphere in Fig. 4(a), the meron texture reliably emerges at the focal plane, albeit with variations in spin density profiles shown in the insets in Fig. 4(g). In this spatially perturbed pupil, near the focal center retains a pure longitudinal spin $s_z$, protected by the intrinsic spin-orbit interaction and the net spin of the input. The boundary structure, in contrast to the uniform pupil case characterized by a purely azimuthal spin $s_\phi$, now consists of a superposition of $s_r$ and $s_\phi$, forming an in-plane spin configuration with negligible longitudinal $s_z$. Despite these modifications, the overall meron topology remains unchanged, demonstrating its intrinsic robustness. Besides random spatial polarization perturbations, we also investigated strong spatial amplitude and phase perturbations to the input pupil, yet the meron structure remains stable, and the details can be found in \cite{supp}.

This meron structure can be verified experimentally using the setup shown in Fig. 5(a). To test the robustness of the meron structure, we superpose two orthogonal CP states incoherently at different intensity ratios $\gamma = I_R/I_L$. This is equivalent to a partially polarized input with a varying degree of polarization $P = |\frac{1-\gamma}{1+\gamma}|$
\cite{eismann2021transverse,lindfors2007local}. The transverse E-fields of the individual CP input at the focal plane of objective MO1 are retrieved using off-axis holography\cite{cuche2000spatial}, and the longitudinal $E_z$ is obtained from the transverse E-field using the Gauss's law\cite{maluenda2021experimental}. From the complex E-fields, the local spin can be obtained\cite{supp}.

The blue dots in Fig. 5(b) show the experimentally measured skyrmion number $N_{sk}$ under varying degrees of polarization $P$. Whenever $P$ is non-zero, $N_{sk}\approx\pm 0.5$ with its sign determined by the dominant CP component, which is in good agreement with the theoretical predictions (dashed lines). In Fig. 5(c-d) and Fig. 5(f-g), we show the detailed spin profiles under partially polarized pupils with $P=0.6$, but are dominated by different CP components. The meron core at the center is clearly visible with a zero $s_\perp$ forming a vortex structure nearby (Fig. 5(c) and (f)) and a nonzero $s_z$ with different signs given by the dominant CP components in the pupil (Fig. 5(d) and (g)). More predominantly, the meron boundary highlighted by a ring of fields with a high 3D degree of polarization $P_{3D}\approx 1$ is clearly visible in Fig. 5(e) and (h). These results confirm the existence of the meron structure under partially polarized focused fields.

In conclusion, we have shown that focused optical fields inherently generate a robust meron spin texture, arising spontaneously from a wide class of input conditions without the need for engineered structures or fine-tuning. Both theoretical prediction and experimental verification confirm that this intrinsic meron structure persists under strong disorder in the input, reflecting an intrinsic stability rooted in phase vortex structures. Our findings identify a naturally occurring and topologically protected spin texture in optics, opening the door to robust topological photonic platforms and enriching the understanding of spin-orbit interaction in structured light.

This work has been supported by the National Key Research and Development Program of China(2021YFA1400700); National Natural Science Foundation of China (Grants No. 62105320); CAS Project for Young Scientists in Basic Research(Grant No.YSBR-049);

\bibliographystyle{apsrev4-2}
\bibliography{reference}

\end{document}